\documentclass[12pt]{article}
\usepackage[a4paper,margin=1in]{geometry}
\usepackage[titletoc,title]{appendix}
\usepackage{changepage}
\usepackage[utf8x]{inputenc}
\usepackage{textcomp,marvosym}
\usepackage{fixltx2e}
\usepackage{amsmath,amssymb}
\usepackage{cite}
\usepackage{nameref,hyperref}
\usepackage{authblk}
\usepackage{algorithm}
\usepackage{algpseudocode}
\usepackage{dsfont}
\usepackage{xcolor}
\usepackage{enumitem}
\usepackage{caption,setspace}
\usepackage{subcaption}
\usepackage{pdflscape}
\usepackage{graphicx}
\usepackage{placeins}
\usepackage{mathtools}

\renewcommand{\eqref}[1]{Equation~(\ref{#1})}

\title{Critical length for the spreading-vanishing dichotomy in higher dimensions}

\author[1]{Matthew~J. Simpson\footnote{To whom correspondence should be addressed. E-mail: matthew.simpson@qut.edu.au}}

\affil[1]{School of Mathematical Sciences, Queensland University of Technology, Brisbane, Queensland 4001, Australia}

\begin{document}

\maketitle
\begin{abstract}
We consider an extension of the classical Fisher-Kolmogorov equation, called the \textit{Fisher-Stefan} model, which is a moving boundary problem on $0 < x < L(t)$.  A key property of the Fisher-Stefan model is the \textit{spreading-vanishing dichotomy}, where solutions with $L(t) > L_{\textrm{c}}$ will eventually spread as $t \to \infty$, whereas solutions where  $L(t) \ngtr L_{\textrm{c}}$ will vanish as $t \to \infty$.  In one dimension is it well-known that the  critical length is $L_{\textrm{c}} = \pi/2$.  In this work we re-formulate the Fisher-Stefan model in higher dimensions and calculate $L_{\textrm{c}}$ as a function of spatial dimensions in a radially symmetric coordinate system.  Our results show how $L_{\textrm{c}}$ depends upon the dimension of the problem and numerical solutions of the governing partial differential equation are consistent with our calculations.
\end{abstract}


\newpage

\section{Introduction}
\label{sec:intro}
The well-known Fisher-Kolmogorov model~\cite{Fisher1937,Kolmogorov1937,Canosa1973,Murray2002,Grindrod2007} is a reaction-diffusion equation that is often used to describe the spatial and temporal spreading of a population density where individuals in that population undergo random diffusive migration and logistic proliferation.  The Fisher-Kolmogorov model is often written as
\begin{equation}\label{eq:Fisher}
\dfrac{\partial u(x,t)}{\partial t} = D \dfrac{\partial ^2 u(x,t)}{\partial x^2} + \lambda u(x,t) \left[ 1 - \dfrac{u(x,t)}{K} \right],
\end{equation}
where $u(x,t)$ is the population density, $x$ is position, $t$ is time, $D$ is the diffusivity, $\lambda$ is the proliferation rate and $K$ is the carrying capacity density.  The Fisher-Kolmogorov equation is often studied on an infinite domain,  $-\infty < x < \infty$,  where it is well known to give rise to travelling wave solutions that eventually propagate with speed $c = 2 \sqrt{D \lambda}$ for initial conditions with compact support~\cite{Fisher1937,Kolmogorov1937,Canosa1973,Murray2002,Grindrod2007}.  The Fisher-Kolmogorov equation, and certain extensions, have been used to model invasion fronts in ecology~\cite{Skellam1951,Shigesada1995,Steel1998,Forbes1997,Broadbridge2002,Bradshaw2004} and cell biology~\cite{Sherratt90,Gatenby1996,Painter2003,Swanson2003,Maini2004a,Maini2004b,Simpson2006,Sengers2007,Simpson2007,Swanson2008,Simpson2013,Treloar2014,Johnston2015,Johnston2016,Jin2016,Nardini2016,Haridas2017,Vittadello2018,Browning2019,Warne2019}.  Despite the widespread application of this fundamental model, there are several shortcomings.  First,  any localised initial condition with compact support will always lead to population growth and successful colonisation.  Therefore, an implicit assumption in the Fisher-Kolmogorov model is that the population will always survive and never go extinct.  Second, solutions of the Fisher-Kolmogorov equation are smooth and without compact support.   This means that the Fisher-Kolmogorov model does not, strictly speaking, lead to a clear and unambiguous invasion front.  This can be problematic because well defined invasion fronts are often observed in practice~\cite{Maini2004a,Maini2004b}.

There are several ways that these two shortcomings of the Fisher-Kolmogorov model have been addressed in the applied mathematics literature.  One common approach is to consider an  extension of Equation (\ref{eq:Fisher})  that incorporates degenerate nonlinear diffusion since this leads to travelling wave solutions  with a well-defined front~\cite{Murray2002,Maini2004a,Maini2004b,Harris2004,Sanchez1994,Sanchez1995,Witsleki1995,Sherratt1996,Simpson2011,Baker2012,McCue2019}.  Another extension that has received less attention in the applied mathematics literature, but far more attention in the analysis literature, is to re-cast Equation (\ref{eq:Fisher}) as a moving boundary problem,
\begin{equation}\label{eq:FisherStefanDimensional}
\dfrac{\partial u(x,t)}{\partial t} = D \dfrac{\partial ^2 u(x,t)}{\partial x^2} + \lambda u(x,t) \left[ 1 - \dfrac{u(x,t)}{K} \right],
\end{equation}
on $0 < x < L(t)$, with $\partial u(0,t) / \partial x = 0$ and $u(L(t),t)=0$, ensuring that there is a well-defined front at the moving boundary, $x=L(t)$.  To close the problem, the evolution of the moving boundary is taken to follow a classical Stefan condition, $\textrm{d}L(t) / \textrm{d}t = -\hat{\kappa} \partial u(L(t),t) / \partial x$~\cite{Crank1987,Gupta2017,McCue2003,McCue2005,McCue2008}.  As we will show, the adoption of Equation (\ref{eq:FisherStefanDimensional}) alleviates both the shortcomings of the classical Fisher-Kolmogorov model since the this moving boundary analogue leads to solutions with well-defined fronts, as well as permitting certain initial conditions to become extinct.

The moving boundary analogue of the Fisher-Kolmogorov model has been called the \textit{Fisher-Stefan} model~\cite{ElHachem2019}.  To simplify our analysis we rescale the variables: $x' = x/\sqrt{D / \lambda}$,  $t'=t \lambda$ and $u'=u/K$. Dropping the prime notation gives
 \begin{equation}\label{eq:FisherStefan}
\dfrac{\partial u(x,t)}{\partial t} = \dfrac{\partial ^2 u(x,t)}{\partial x^2} + u(x,t) \left[ 1 - u(x,t) \right],
\end{equation}
on $0 < x < L(t)$, with $\partial u(0,t) / \partial x = 0$, $u(L(t),t)=0$ with $\textrm{d}L(t) / \textrm{d}t = -\kappa \partial u(L(t),t) / \partial x$. In this nondimensional framework there is just one parameter, $\kappa > 0$.  Varying $\kappa$ allows us to control the finite speed of the moving boundary at $x=L(t)$.   This mathematical model has been studied extensively in the analysis literature~\cite{Du2010,Du2011,Bunting2012,Du2012,Du2014a,Du2014b,Du2015}.  Much of this analysis has centred upon proving properties of travelling wave solutions of Equation (\ref{eq:FisherStefan}) as well as characterising the \textit{spreading-vanishing dichotomy}. As we will show, it is very interesting that certain initial conditions and choices of $\kappa$ in Equation (\ref{eq:FisherStefan}) lead to eventual \textit{spreading} in the form of a travelling wave, whereas other choices of initial condition and $\kappa$ in Equation (\ref{eq:FisherStefan}) lead to \textit{extinction}.  A key feature of the spreading-vanishing dichotomy is that there exists a critical length, $L_{\textrm{c}}$, such that solutions of Equation (\ref{eq:FisherStefan}) that evolve in such a way that $L(t) > L_{\textrm{c}}$ always lead to eventual  spreading, whereas solutions that evolve in such a way so that $L(t) \ngtr L_{\textrm{c}}$ always leads to eventual extinction.  A great deal of attention has been paid to the analysis of this spreading-vanishing dichotomy, and establishing that $L_{\textrm{c}} = \pi /2$~\cite{Du2010,Du2011,Bunting2012,Du2012,Du2014a,Du2014b,Du2015}.

The aim of this work is to revisit the spreading-vanishing dichotomy in a more general setting by re-casting Equation (\ref{eq:FisherStefan}) in a radially symmetric geometry.  Using numerical simulations to guide our calculations, we provide a very straightforward interpretation of $L_{\textrm{c}}$, and show how the critical length depends upon the dimension of the problem.  All results are supported by numerical evidence.  The details of the numerical scheme we use to solve Equation (\ref{eq:FisherStefan}) is described in the Appendix, and MATLAB software to implement the numerical scheme and to explore and visualise various solutions is available on \href{https://github.com/ProfMJSimpson/SVD_2020}{GitHub}.

\section{Results and Discussion}
\label{sec:results}
We will first revisit the spreading-vanishing dichotomy in a one-dimensional coordinate system before moving on to show that the calculations can be extended to higher dimensions.

\subsection{Preliminary results in one dimension}
\begin{figure}
	\centering
	\includegraphics[width=0.95\linewidth]{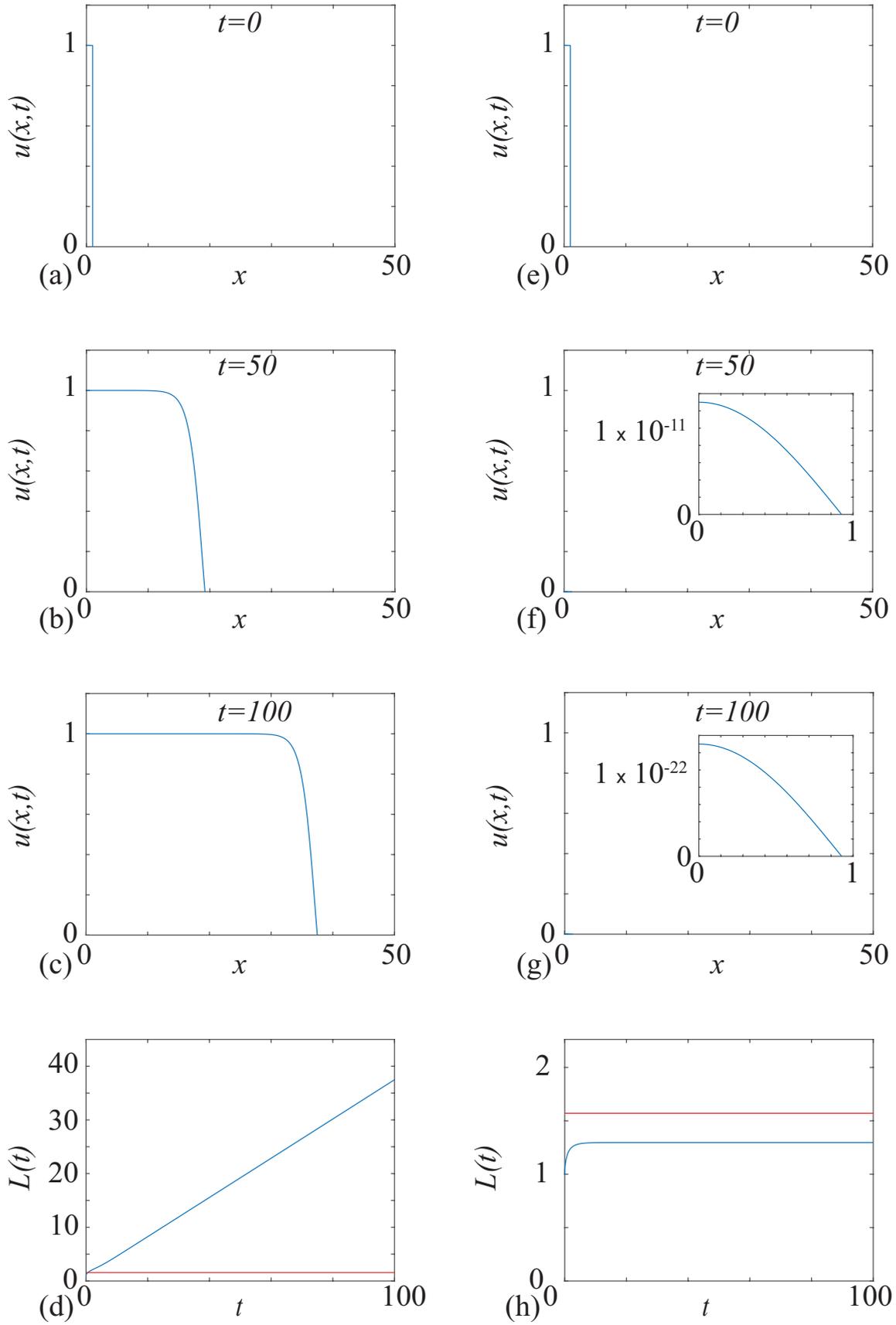}
	\caption{Numerical solutions of Equation (\ref{eq:FisherStefan}).  Results in (a)-(d) are for $\kappa=1$ and lead to a spreading travelling wave. Results in (e)-(h) are for $\kappa = 0.2$  and lead to extinction. (d) and (h) show the evolution of $L(t)$ (blue) and $L_{\textrm{c}} = \pi/2$ (red).}
	\label{fig:preliminaryresults}
\end{figure}

We first begin by visualising numerical solutions of Equation (\ref{eq:FisherStefan}).  For simplicity, all numerical solutions in this work have the same initial condition: $u(x,0)=1$ for $0 \le x \le 1$ with $L(0)=1$.  The solution of Equation (\ref{eq:FisherStefan}) with $\kappa=1$ is given in Figure \ref{fig:preliminaryresults}(a)-(d) showing that the initial density evolves into a constant speed, constant shape travelling wave.  Insights into such travelling waves can be obtained by phase plane analysis~\cite{ElHachem2019}.  The evolution of $L(t)$ is given in Figure \ref{fig:preliminaryresults}(d) where we see that $L(t)$ eventually increases linearly as $t \to \infty$.

With $\kappa=0.2$ the solution of Equation (\ref{eq:FisherStefan}) is given in Figure \ref{fig:preliminaryresults}(e)-(h).  These solutions show that the initial density evolves in such a way that the population goes extinct as $t \to \infty$.   Solutions at $t=50$ and $t=100$ (Figure \ref{fig:preliminaryresults}(f)-(g)) are plotted with an inset because these solutions are so close to zero that it is impossible to clearly visualise the solutions when they are plotted on the same scale as the other solutions in Figure \ref{fig:preliminaryresults}.  The evolution of $L(t)$ in Figure \ref{fig:preliminaryresults}(h) shows that $L(t)$ initially increases, and then asymptotes to some constant value as $t \to \infty$.  In this case we always have $\partial u(x,t) /\partial x < 0$ at $L(t)$, and so the Stefan condition ensures that $\textrm{d}L(t)/\textrm{d}t \ge 0$.  This means that $L(t)$ never decreases.

This simple numerical demonstration in Figure \ref{fig:preliminaryresults} illustrates the key features of the spreading-vanishing dichotomy.  Unlike the usual Fisher-Kolmogorov model where all positive initial conditions will eventually lead to spreading, here the moving boundary analogue of this model can either lead to an invading travelling wave or it can lead to extinction. Numerical solutions of Equation (\ref{eq:FisherStefanDimensional}) in Figure \ref{fig:preliminaryresults} provide some intuition about how we may  analyse this dichotomy.  As illustrated in Figure \ref{fig:preliminaryresults}, When extinction occurs we always have $u(x,t) \to 0$ as $t \to \infty$ for all $0 \le x \le L(t)$, suggesting that we can analyse the behaviour in this regime by  linearising.  If $v(x,t) \ll 1$, an appropriate linear analogue of the Fisher-Stefan model is
\begin{equation}\label{eq:FisherStefanlinearizeded}
\dfrac{\partial v(x,t)}{\partial t} = \dfrac{\partial ^2 v(x,t)}{\partial x^2} + v(x,t),
\end{equation}
on $0  < x < L(t)$ with $\partial v(0,t) / \partial x = 0$ and $v(L(t),t)=0$.  For the purposes of this analysis we treat the domain length as fixed, $L(t) = L$. The exact solution of Equation (\ref{eq:FisherStefanlinearizeded}) is
\begin{equation}\label{eq:FisherStefanlinearizededsolution}
v(x,t) = \sum_{n=1}^{\infty} A_n \cos(\lambda_n x)\textrm{e}^{-t(\lambda_n^2-1)},
\end{equation}
where $\lambda_n = \pi(2n-1)/(2L)$ for $n=1,2,3, \ldots$, and the coefficients $A_n$ can be chosen so that the solution matches an initial condition.  As we will show, our analysis does not require that we specify these coefficients.   We now approximate Equation (\ref{eq:FisherStefanlinearizededsolution}) by retaining just the first term in the series.  Such leading eigenvalue approximations can be particularly accurate as $t \to \infty$~\cite{Hickson2009a,Hickson2009b,Simpson2015},
\begin{equation}\label{eq:leadingeigenvalue1D}
V(x,t) \sim A_1 \cos\left(\dfrac{\pi x}{2L}\right)\textrm{e}^{-t\left[\left(\dfrac{\pi}{2L}\right)^2-1\right]}.
\end{equation}
With this solution we can specify a conservation statement for the time evolution of the total population in the domain,
\begin{equation}\label{eq:conservation1D}
\dfrac{\textrm{d} m}{\textrm{d} t} = \underbrace{\int_{0}^{L}V(x,t) \, \textrm{d}x}_{\textrm{accumulation due to source term}} - \underbrace{-\dfrac{\partial V(L,t)}{\partial x}}_{\textrm{loss due to the diffusive flux at $x=L$}}.
\end{equation}
Setting $\textrm{d} m / \textrm{d}t = 0$ we solve Equation (\ref{eq:conservation1D}) using $V(x,t)$, giving $L_\textrm{c} = \pi / 2 \approx 1.570796327$, which is independent of $A_n$.  This means that whenever a solution of Equation (\ref{eq:FisherStefan}) evolves in such a way that $L(t) > \pi/2$, the accumulation due to the source term is greater than the loss due to the diffusive flux at the moving boundary. In contrast, whenever $L(t) < \pi/2$, the accumulation due to the source term is less than the loss at the moving boundary.  This result has some interesting consequences and provides a straightforward explanation for our numerical results in Figure \ref{fig:preliminaryresults}.  For our initial condition we have $L(0) = 1 < \pi/2$ and it is unclear whether or not the transient solution will evolve such that $L(t) >  \pi/2$ as $t \to \infty$.  When $\kappa$ is sufficiently large we see that the evolution of the solution is such that $L(t)$ eventually exceeds $\pi/2$ and at this point the population will continue to grow indefinitely, as shown in Figure \ref{fig:preliminaryresults}(d).  In contrast, for sufficiently small $\kappa$, the evolution of the solution is such that $L(t)$ that never reaches $\pi/2$ and so the diffusive flux out of the domain is greater than the production due to the source term and we see eventual extinction where $\displaystyle{\lim_{t \to \infty}L(t) < \pi/2}$, as shown in Figure \ref{fig:preliminaryresults}(h).

Our numerical results and analysis of the linearised problem confirm that we have $L_{\textrm{c}} = \pi/2$ in one dimension and this explains our original choice of initial condition in Figure \ref{fig:preliminaryresults} where $L(0)=1 < L_{\textrm{c}}$.  In this case $L(0) < L_{\textrm{c}}$ and without solving for the time dependent solution it is unclear whether $L(t)$ evolves in such a way that it ever exceeds $L_{\textrm{c}}$.  Therefore, with this initial condition we have the flexibility of choosing $\kappa$ to be sufficiently small such that $L(t)$ never reaches $L_{\textrm{c}}$ and the population goes extinct, or we can choose $\kappa$ to be sufficiently large that $L(t)$ will eventually exceed $L_{\textrm{c}}$ and lead to a travelling wave.  Had we chosen a different initial condition with $L(0) > L_{\textrm{c}}$ we would always observe the formation of a travelling wave provided $\kappa >0$.  Since we are interested in studying the spreading-extinction dichotomy we restrict our numerical solutions to the case where $L(0) < L_{\textrm{c}}$. Setting $L(0)=1$ is a convenient way to achieve this.

While it has long been known that $L_\textrm{c} = \pi/2$ in the more formal literature~\cite{Du2010,Du2011,Bunting2012,Du2012,Du2014a,Du2014b,Du2015}, here we offer a very simple and intuitive way to calculate and interpret the spreading-vanishing dichotomy in terms of this critical length.  Our approach is mathematically and conceptually straightforward, and as we will now show, also applies in other geometries.

\subsection{Critical length in higher dimensions}
A generalisation of Equation (\ref{eq:FisherStefan}) is to consider the Fisher-Stefan model in a radially symmetric coordinate system,
\begin{equation}\label{eq:FisherStefan_ndimensions}
\dfrac{\partial u(x,t)}{\partial t} = \dfrac{\partial^2 u(x,t)}{\partial x^2} + \dfrac{d-1}{x}\dfrac{\partial u(x,t)}{\partial x} + u(x,t)\left[ 1- u(x,t) \right],
\end{equation}
for $t > 0$ and $0 < x < L(t)$, with $\partial u(0,t)/\partial x = 0$,  $u(L(t),t)=0$, and $\textrm{d} L(t) / \textrm{d} t = -\kappa \partial u(L(t),t) / \partial x$.  Here $d=1,2,3$ is the dimension.  We will now explore how $L_{\textrm{c}}$ depends upon $d$.

\begin{figure}[htp]
	\centering
	\includegraphics[width=0.95\linewidth]{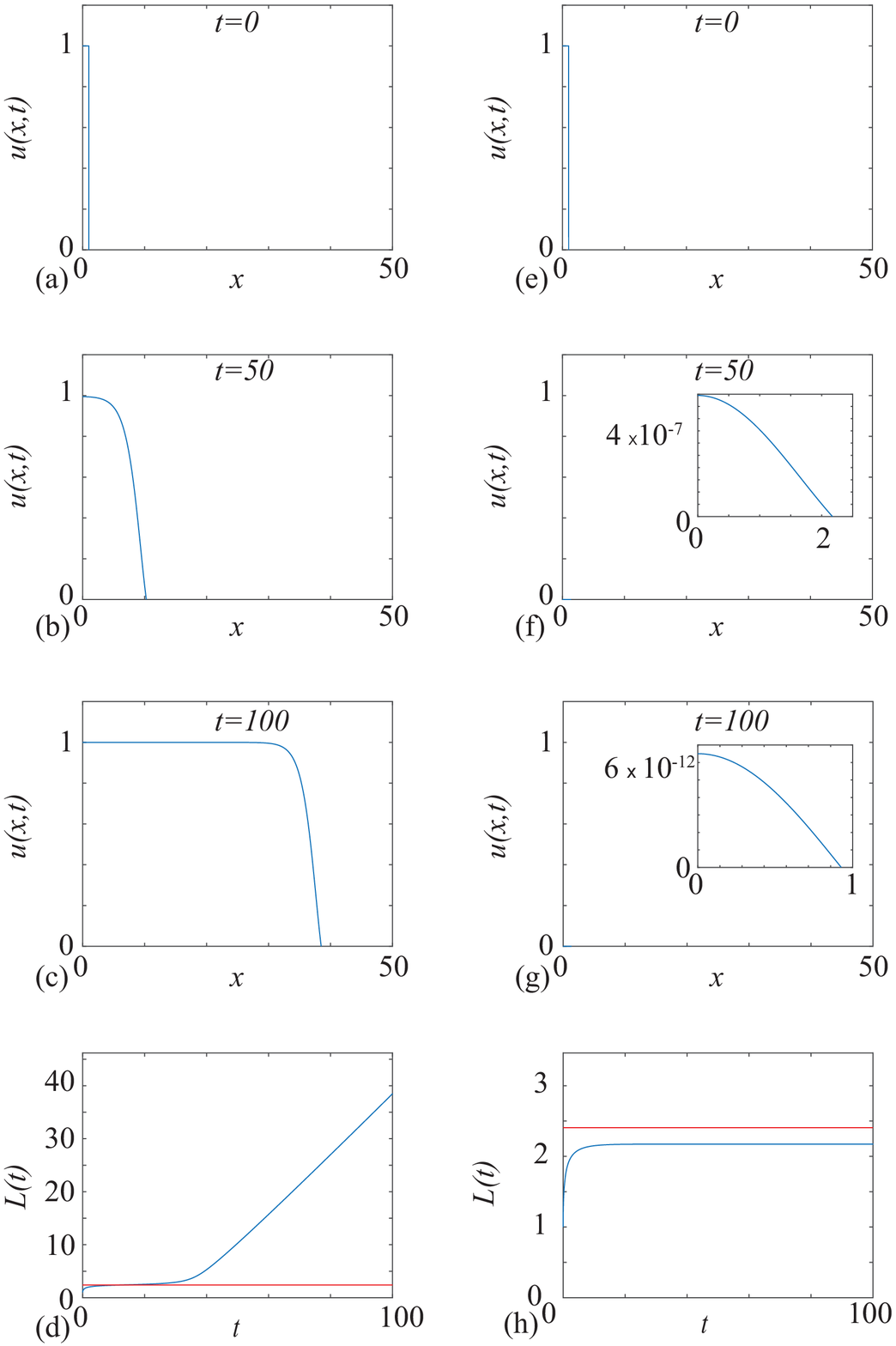}
\caption{Numerical solutions of Equation (\ref{eq:FisherStefan_ndimensions}) with $d=2$.  Results in (a)-(d) show a solution with $\kappa2.3$ leading to spreading. Results in (e)-(h) show a solution with $\kappa = 2.1$ leading to extinction. (d) and (h) show the evolution of $L(t)$ (blue) and $L_{\textrm{c}} = 2.404825558$ (red).}
	\label{fig:disc}
\end{figure}

Setting $d=2$ in Equation (\ref{eq:FisherStefan_ndimensions}) allows us to consider the spreading-vanishing dichotomy on a disc.  Preliminary numerical simulations indicate that for a particular choice of $u(x,0)$ the solution may either evolve into a moving front when $\kappa$ is sufficiently large, or the solution may go extinct when $\kappa$ is sufficiently small, just like we demonstrated in Figure \ref{fig:preliminaryresults} when $d=1$.  To explore this we linearise, with $v(x,t) \ll 1$, to obtain
\begin{equation}\label{eq:linearizededdisc}
\dfrac{\partial v(x,t)}{\partial t} = \dfrac{\partial ^2 v(x,t)}{\partial x^2} + \dfrac{1}{x} \dfrac{\partial v(x,t)}{\partial x} + v(x,t),
\end{equation}
on $0  < x < L$, with $\partial v(0,t) / \partial x = 0$ and $v(L,t)=0$.  The exact solution of Equation (\ref{eq:linearizededdisc}) is
\begin{equation}\label{eq:discsolution}
v(x,t) = \sum_{n=1}^{\infty} A_n J_0 \left( x \sqrt{1 + \lambda_n^2} \right) \textrm{e}^{-\lambda_n^2 t},
\end{equation}
where $J_0(x)$ is the zeroth-order Bessel function of the first kind.  Here the eigenvalues, $\lambda_n$, are obtained by setting $L\sqrt{1 + \lambda_n^2}$ equal to the $n^{\textrm{th}}$ zeros of $J_0(x)$. If we invoke a leading eigenvalue approximation we have $\sqrt{1 + \lambda_1^2} = z/L$, where $z$ is the first zero of $J_0(x)$, or $z \approx 2.404825558$.  The leading eigenvalue approximation is
\begin{equation}\label{eq:discsolution1term}
V(x,t) \sim A_1 J_0\left(\dfrac{zx}{L}\right)\textrm{e}^{-t\left[\left(\dfrac{z}{L}\right)^2 -1\right]}.
\end{equation}
To proceed we formulate a condition where the accumulation due to the source term is precisely matched by the loss at the moving boundary. Taking care to account for the geometry we obtain,
\begin{equation}\label{eq:conservation}
\underbrace{\int_{0}^{L}V(x,t) 2 \pi x \, \textrm{d}x}_{\textrm{accumulation due to source term}} = \underbrace{-\dfrac{\partial V(L,t)}{\partial x}2 \pi L}_{\textrm{loss due to the diffusive flux at $x=L$}},
\end{equation}
in which we can substitute Equation (\ref{eq:discsolution1term}) and solve for $L$.  This procedure shows that $L_{\textrm{c}}$ is the first zero of $J_0(x)$, giving $L_{\textrm{c}} \approx 2.404825558$.  Numerical solutions in Figure \ref{fig:disc} confirm this.

\begin{figure}[htp]
	\centering
	\includegraphics[width=0.95\linewidth]{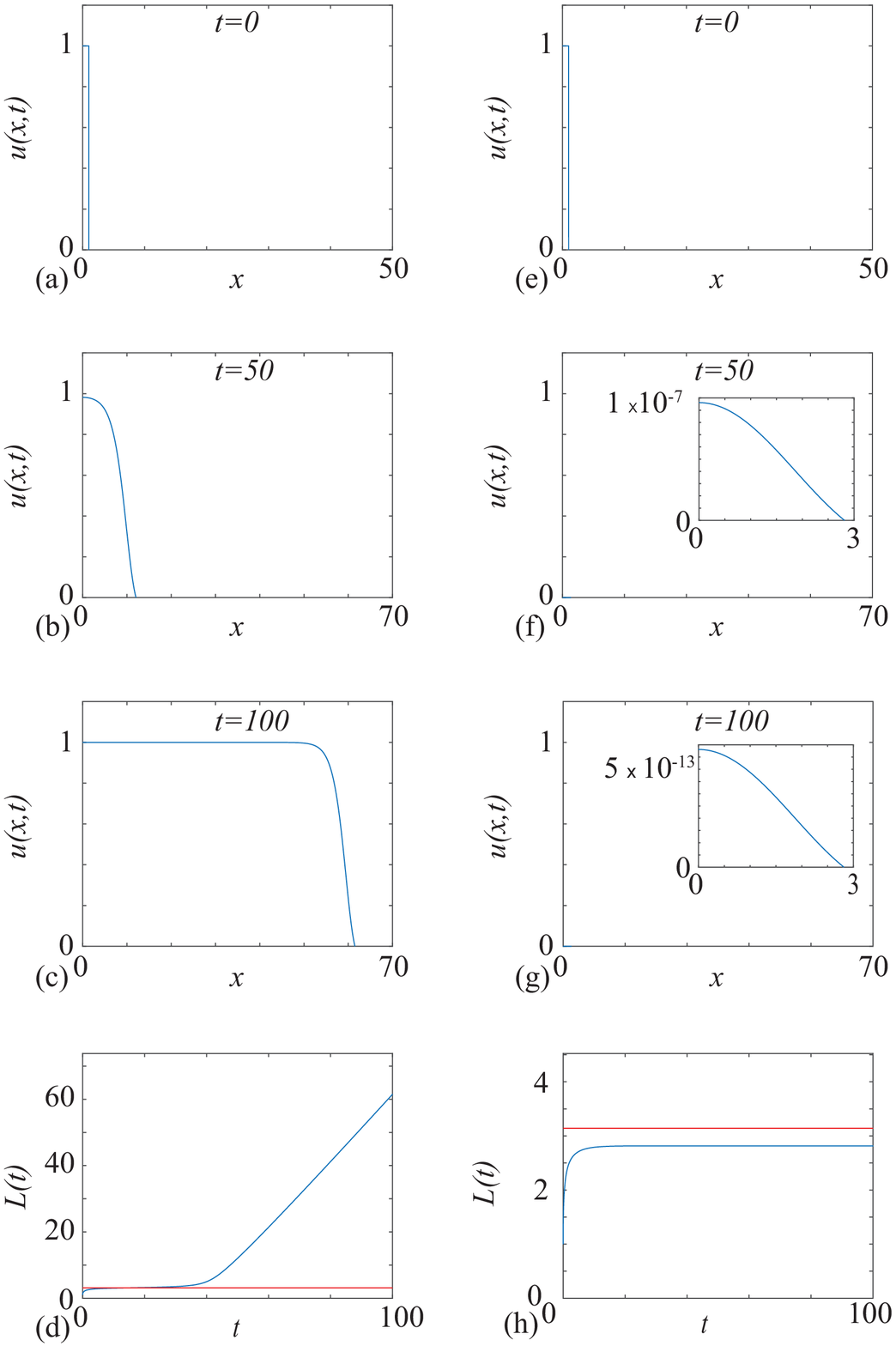}
	\caption{Numerical solutions of Equation (\ref{eq:FisherStefan_ndimensions}) with $d=3$.  Results in (a)-(d) show a solution with  $\kappa= 11$ leading to spreading. Results in (e)-(h) show a solution with $\kappa = 10$ leading to extinction. (d) and (h) show the evolution of $L(t)$ (blue) and $L_{\textrm{c}} = \pi$ (red).}
	\label{fig:sphere}
\end{figure}

Setting $d=3$ in Equation (\ref{eq:FisherStefan_ndimensions}) allows us to consider the spreading-vanishing dichotomy on a sphere.  To explore the critical length in this case we linearise, with $v(x,t) \ll 1$, to obtain
\begin{equation}\label{eq:linearizededsphere}
\dfrac{\partial v(x,t)}{\partial t} = \dfrac{\partial ^2 v(x,t)}{\partial x^2} + \dfrac{2}{x} \dfrac{\partial v(x,t)}{\partial x} + v(x,t),
\end{equation}
on $0  < x < L$, with $\partial v(0,t) / \partial x = 0$ and $v(L,t)=0$.  The exact solution of Equation (\ref{eq:linearizededsphere}) is
\begin{equation}\label{eq:spheresolution}
v(x,t) = \sum_{n=1}^{\infty} \dfrac{A_n}{x} \sin \left( x \sqrt{1 + \lambda_n^2} \right) \textrm{e}^{-\lambda_n^2 t},
\end{equation}
where $\lambda_n = n\pi / L$, for $n=1,2,3, \ldots$.  The associated leading eigenvalue approximation is
\begin{equation}\label{eq:spheresolution1term}
V(x,t) \sim \dfrac{A_1}{x} \sin \left[ \dfrac{\pi x}{L} \right]\textrm{e}^{-t\left[\left(\dfrac{\pi}{L}\right)^2-1\right]}.
\end{equation}
The associated conservation statement is,
\begin{equation}\label{eq:conservation}
\underbrace{\int_{0}^{L}V(x,t) 4 \pi x^2 \, \textrm{d}x}_{\textrm{accumulation due to source term}} = \underbrace{-\dfrac{\partial V(L,t)}{\partial x}4 \pi L^2}_{\textrm{loss due to the diffusive flux at $x=L$}},
\end{equation}
in which we substitute Equation (\ref{eq:spheresolution1term}) and solve for $L$, giving $L_{\textrm{c}} = \pi \approx 3.141592654$ Numerical solutions in Figure \ref{fig:sphere} confirm this.

\section{Conclusions}
\label{sec:discuss}
In this work we explore an extension of the well-known Fisher-Kolmogorov model, called the Fisher-Stefan model.  This extension involves reformulating the Fisher-Kolmogorov model as a moving boundary problem, and the evolution of the moving boundary is described by a classical Stefan condition.  The Fisher-Stefan model has two key advantages over the usual Fisher-Kolmogorov model: (i) solutions of the Fisher-Stefan model are characterised by a sharp front, which is often relevant in practice; and, (ii) solutions of the Fisher-Stefan model do not always lead to successful colonisation since the population can go extinct as $t \to \infty$.  An important feature of the Fisher-Stefan model is the existence of a critical length, $L_{\textrm{c}}$.  As we show, solutions of the Fisher-Stefan model that evolve in such a way that $L(t) > L_{\textrm{c}}$ always lead to spreading, whereas solutions that evolve in such a way so that $L(t) \ngtr L_{\textrm{c}}$ always leads to extinction.  Formal analysis of the spreading-vanishing dichotomy has established that $L_{\textrm{c}} = \pi /2$ in a standard one-dimensional Cartesian geometry~\cite{Du2010,Du2011,Bunting2012,Du2012,Du2014a,Du2014b,Du2015}. This property is often called the spreading-vanishing dichotomy, and it has been extensively studied in one-dimension.  Far less attention has been paid to similar behaviour in higher dimensions.

In this work we re-formulate the Fisher-Stefan model in a radially symmetric coordinate system and study the spreading-vanishing dichotomy in one, two and three dimensions.  Using simple conservation arguments we confirm that   $L_{\textrm{c}} = \pi /2$ in the usual one-dimensional geometry, and this result is confirmed by numerical solutions.  By extending these ideas to consider the Fisher-Stefan model on a disc and sphere we find that the critical length depends upon the geometry of the problem.   In particular the critical length for the Fisher-Stefan model on a disc is the first zero of the zeroth-order Bessel function of the first kind ($L_{\textrm{c}} \approx 2.404825558$) and the critical length for the Fisher-Stefan model on a sphere is $\pi$ ($L_{\textrm{c}} \approx 3.141592654$).  Numerical solutions of the governing partial differential equation are consistent with these findings.

Our approach for calculating the critical length is conceptually straightforward,  and shows how the critical length depends upon the dimension.  A great deal of literature about the spreading-vanishing dichotomy involves rigorous analysis that can be both difficult to interpret and difficult to extend to more general problems.  Our approach, in contrast, is quite flexible and there are many potential extensions.  While we have not pursed further calculations for other choices of $d$, it is straightforward to apply the methods described here for $d > 3$ to explore the spreading-vanishing dichotomy on an hypersphere.  For example, setting $d=4$ in Equation (\ref{eq:FisherStefan_ndimensions}) and repeating the analysis shows that the critical length is the first zero of the first-order Bessel function of the first kind ($L_{\textrm{c}} \approx 3.831705970$).  Further calculations for different $d$ can also be performed.

We close by commenting on some extensions of Equation (\ref{eq:FisherStefan_ndimensions}) that could be of interest.  As we already pointed out in the Introduction, a very common extension of the classical Fisher-Kolmogorov model is to generalise the linear diffusion term to a degenerate nonlinear diffusion term~~\cite{Murray2002,Maini2004a,Maini2004b,Harris2004,Sanchez1994,Sanchez1995,Witsleki1995,Sherratt1996,Simpson2011,Baker2012,McCue2019}.  This generalisation is often motivated by desire to seek solutions with a well-defined sharp front~\cite{Murray2002,Maini2004a,Maini2004b,Harris2004,Sanchez1994,Sanchez1995,Witsleki1995,Sherratt1996,Simpson2011,Baker2012,McCue2019}.  While it is certainly possible generalise Equation (\ref{eq:FisherStefan_ndimensions}) to include degenerate nonlinear diffusion~\cite{Fadai2020}, an important consequence of this in the moving boundary context is that this extension does not lead to a spreading-vanishing dichotomy.  This is because the flux at the moving boundary $x=L(t)$, where $u(L(t),t)=0$, is always zero thereby always preventing extinction.  In contrast, the one-phase Stefan condition in  Equation (\ref{eq:FisherStefan_ndimensions}) leads to a loss of mass at the moving boundary and it is this loss of mass that enables the population to go extinct.   Another interesting extension would be to consider the spreading-vanishing dichotomy in higher dimensions for non radial geometries and explore the relationship between the solution in radial and non-radial geometries~\cite{Vazquez2007,Barenblatt1996}.  Such an extension to consider moving boundary problems in higher dimensions without radial symmetry would require a very different approach numerically and so we leave this for future consideration.

\newpage

\section{Appendix: Numerical Methods}
\label{sec:appendix}
We develop a simple, robust numerical solution of
\begin{equation}\label{eq:pdenumericalmethod}
\dfrac{\partial u(x,t)}{\partial t} = \dfrac{\partial^2 u(x,t)}{\partial x^2} + \dfrac{d-1}{x}\dfrac{\partial u(x,t)}{\partial x} + u(x,t)\left[ 1- u(x,t) \right],
\end{equation}
for $t > 0$ and $0 < x < L(t)$, with $u(L(t),t)=0$, $\partial u(0,t)/\partial x = 0$ and $\textrm{d} L(t) / \textrm{d} t = -\kappa \partial u(L(t),t) / \partial x$.  To solve this partial differential equation we first re-cast the moving boundary problem on a fixed domain by setting $\xi = x / L(t)$, giving
\begin{equation}\label{eq:pdetransformed}
\dfrac{\partial u(\xi,t)}{\partial t} = \dfrac{1}{L^2(t)}\dfrac{\partial^2 u(\xi,t)}{\partial \xi^2} + \left[\dfrac{\xi}{L(t)}\dfrac{\textrm{d}L(t)}{\textrm{d}t} + \dfrac{d-1}{\xi L^2(t)} \right]\dfrac{\partial u(\xi,t)}{\partial \xi} + u(\xi,t)\left[ 1- u(\xi,t) \right],
\end{equation}
for $t > 0$ and $0 < \xi < 1$, with $u(1,t)=0$, $\partial u(0,t) / \partial \xi=0$ and $\textrm{d} L(t) / \textrm{d} t = -(\kappa/L(t)) \partial u(1,t) / \partial \xi$.

We discretise Equation (\ref{eq:pdetransformed}) on a uniform mesh with spacing $\Delta \xi$, and the value of $u(\xi,t)$ at each node is indexed with a subscript $i=1,2,3,\ldots, N$, where $N = 1/\Delta \xi + 1$.  Integrating through time using an implicit Euler approximation, using a subscript $m$ to denote the time step, leads to the following system of $N$ nonlinear algebraic equations,
\begin{align}\label{eq:discretized}
u_1^m &= u_2^m, \notag \\
\dfrac{u_i^{m} -u_i^{m-1}}{\Delta t} &= \dfrac{\left(u_{i-1}^m -2u_{i}^{m} + u_{i+1}^m \right)}{(\Delta \xi L^m)^2} + \left[\dfrac{\xi_i}{L^m}\dfrac{\textrm{d}L^m}{\textrm{d}t} + \dfrac{d-1}{\xi_i (L^m)^2} \right] \dfrac{\left(u_{i+1}^m - u_{i+1}^m \right)}{2 \Delta \xi } \notag \\
& + u_i^{m}\left[1- u_i^{m} \right], \quad i=2,3,4,\ldots N-1, \notag \\
u_{N}^m &= 0,
\end{align}
where $\textrm{d}L^m / \textrm{d}t = \kappa u_{N-1}^m / (L^m \Delta \xi)$. This system of nonlinear algebraic equations is solved using Newton-Raphson iteration until the change in estimates of $u_i^m$ per iteration fall below a tolerance of $\varepsilon$.  This algorithm provides estimates of $u(\xi,t)$ on $ 0 < \xi < 1$ which are then re-scaled to give $u(x,t)$ on $0 < x < L(t)$.  A MATLAB implementation of this algorithm is available on \href{https://github.com/ProfMJSimpson/SVD_2020}{GitHub}.  All results presented in this work use $\Delta \xi = 1 \times 10^{-3}$,  $\Delta t = 1 \times 10^{-4}$ and $\varepsilon = 1 \times 10^{-10}$.  This choice of discretisation appears to be suitable for our purposes, but numerical results should always be checked to test that they are independent of the mesh.  Experience suggests that problems with larger $\kappa$ require particular care to ensure grid convergence.

\paragraph{Acknowledgements} This work was supported by the Australian Research Council (DP170100474).  I appreciate many informal discussions on this topic with Scott McCue, Yihong Du, Wang Jin and Maud El-Hachem, together with helpful suggestions of three anonymous referees.

\end{document}